\documentclass[a4paper,11pt]{article}
\usepackage{pos}

\title{Chiral Dynamics: Theory and Experiment -- A Tribute to Aron Bernstein}

\author*[a,b,c]{Ulf-G. Mei{\ss}ner}

\affiliation[a]{Helmholtz-Institut f\"ur Strahlen- und Kernphysik and Bethe
        Center for Theoretical Physics,
        Universit\"at Bonn,  D-53115 Bonn, Germany} 

\affiliation[b]{Institute for Advanced Simulation, Institut f\"{u}r Kernphysik
        and J\"ulich Center for Hadron Physics, Forschungszentrum J\"{u}lich,
        D-52425 J\"{u}lich, Germany}

\affiliation[c]{Tbilisi State  University,  0186 Tbilisi, Georgia}

\emailAdd{meissner@hiskp.uni-bonn.de}

\abstract{I review and discuss the contributions of Aron Bernstein to the field of chiral dynamics.}

\FullConference{%
  The 10th International Workshop on Chiral Dynamics - CD2021\\
  15-19 November 2021\\
  Online
  }

\begin{document}
\maketitle

\section{Introductory remarks}

\vspace{-2mm}

This talk is about Aron Bernstein and the footprints he left in the field of chiral
dynamics over more than three decades. It is a personal recollection that intends to give
a flavor about the works of a brilliant physicist and great human being. In fact,
I came across Aron's name first in the late 1970ties, when I was an undergraduate at Bochum
studying QFT  using  Schweber's book \cite{Schweber} and noticed the paragraph on the experimental
verification of Delbr\"uck scattering on page~595~\cite{BM}. Later, we became very good friends
and had numerous discussions on physics and other topics. I dearly miss him.

\begin{figure}[b!]
\includegraphics[width=0.45\textwidth]{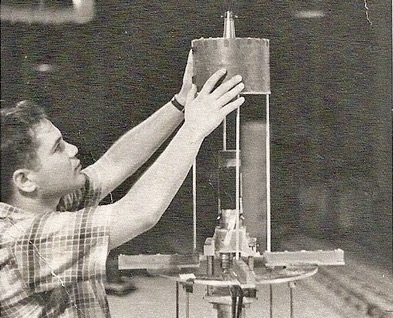}~~
\includegraphics[width=0.54\textwidth]{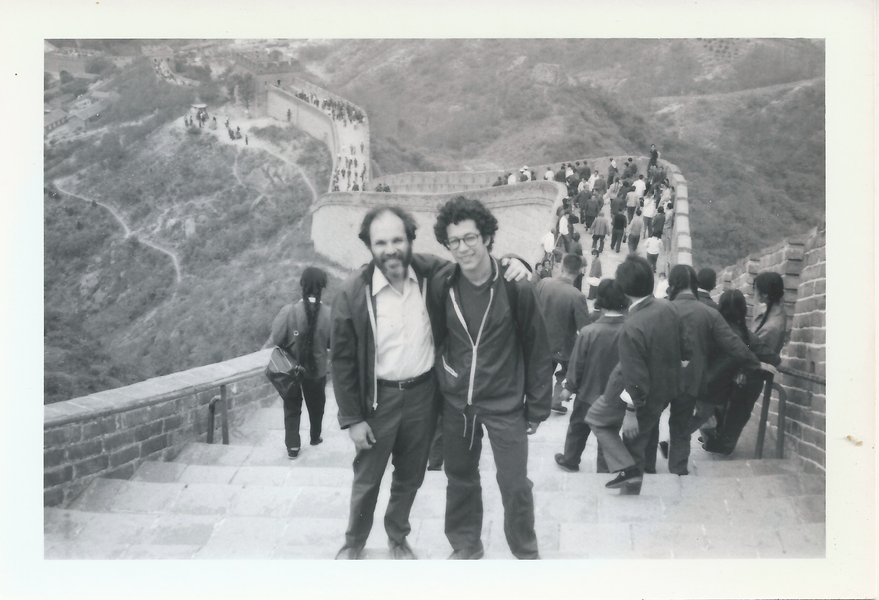}~~
\caption{Left panel: Aron during this postdoc time at Princeton.
Right panel: Aron with his son on the Great Wall in the late 1970ties.
Pictures courtesy of the Bernstein family.
}  
\label{fig:Aron}
\vspace{-0mm}
\end{figure}

Let me start with a short CV of Aron. He was born April 6, 1931, and grew up in Brooklyn
and Queens. In 1953, he obtained his B.Sc. in physics at Union  College in Schenectady, New York,
and 5 years later, he was awarded the PhD in physics from the University  of Pennsylvania,
working on Delbr\"uck scattering. He went on as a postdoc to Princeton, where he started
nuclear physics (NP) research, see the left panel of Fig.~\ref{fig:Aron}, and became an assistant professor at MIT in 1961, performing numerous low-energy NP investigations at the then operating Markle Cyclotron. With the support of Victor Weisskopf, he was promoted to associate professor
in 1966 and became full professor in 1975. To my recollection, Aron became interested in low-energy
QCD and chiral dynamics in the late 1980ties/early 1990ties. Together with Barry Holstein, he
initiated the ``Chiral Dynamics - Theory and Experiment'' workshop series that started in 1994
at MIT, with a rather unusual format. Besides the plenary talks delivered by world-leading
experts in chiral dynamics, there were true working groups where people sat together and discussed the
interplay of theory and experiment to advance the field~\cite{Bernstein:1995vvb}.
The second meeting took place in Mainz in 1997, followed by Jefferson Lab in 2000, Bonn
in 2003 until this, the tenth meeting of this series in 2021 in Beijing (online).
Aron spent many months in Mainz working at the MAMI accelerator on issues related to
threshold pion photoproduction, see also Sec.~\ref{sec:photo}, supported by a prestigious
Humboldt research prize, and later was one of the initiators of the PrimEx experiment
at Jefferson Lab to measure precisely the neutral pion lifetime and test the chiral
anomaly of QCD, see also  Sec.~\ref{sec:anom}. Besides all his works in physics, which
I can not possibly cover, he was a life-long arms control activist, a topic that was always
very dear to his heart. Aron passed away on January 14, 2020, leaving an impressive legacy
which I can only recount in small parts in the following.

\section{Threshold pion photoproduction off nucleons}
\label{sec:photo}

\begin{figure}[t]
\begin{center}  
\includegraphics[trim=0 -15ex 0 0,width=0.35\textwidth]{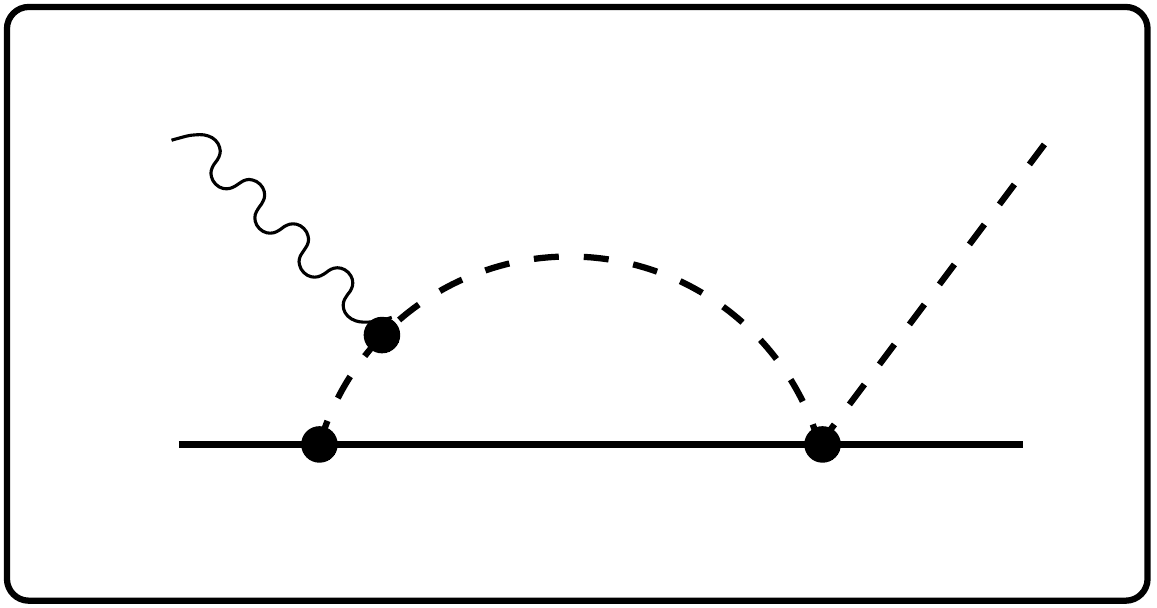}~~~~~~~~~~~~
\includegraphics[width=0.40\textwidth]{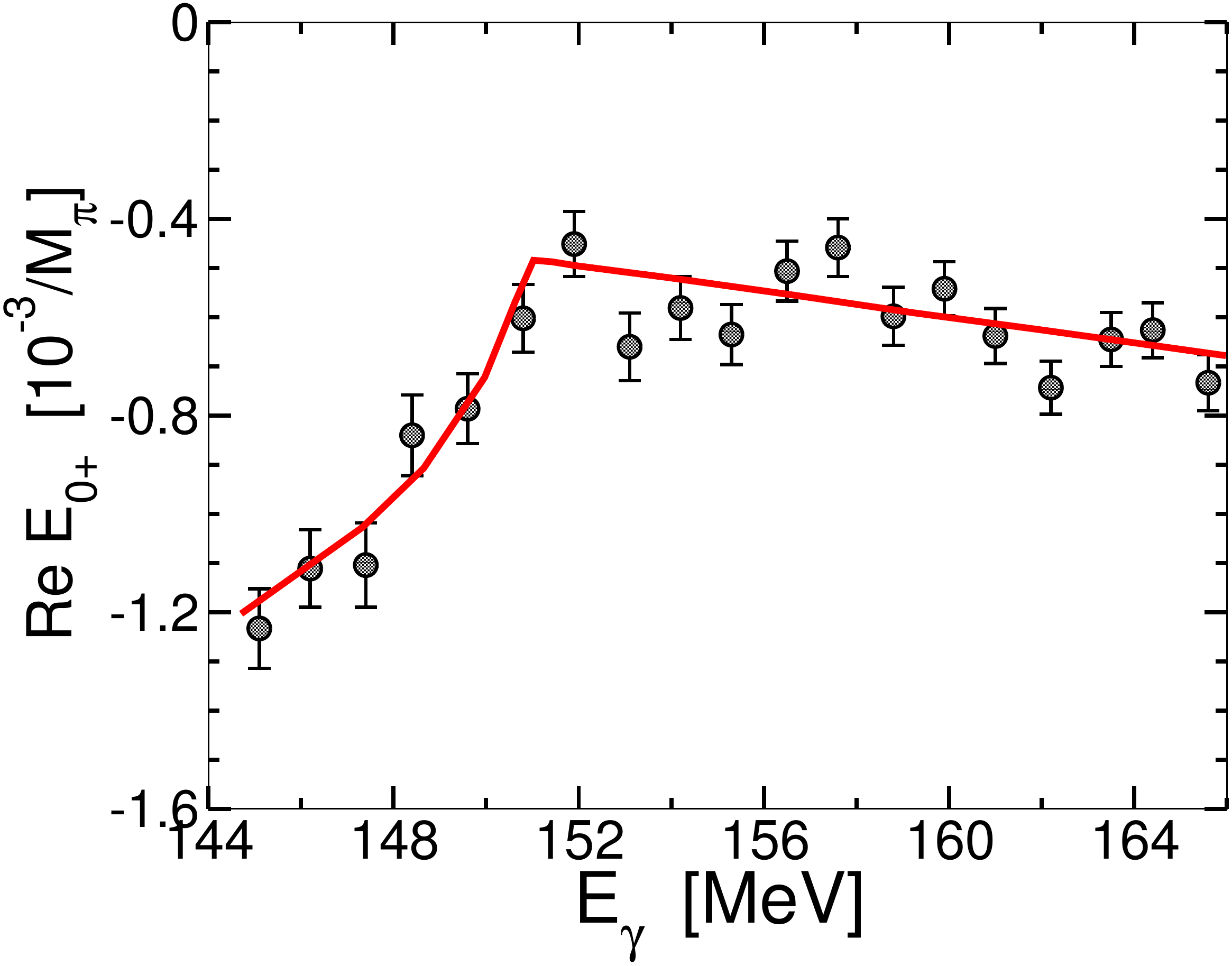}
\end{center}
\vspace{-5mm}
\caption{Left panel: The triangle diagram in neutral pion photoproduction. Solid, dashed and wiggly lines
denote nucleons, pions and photons, in order. The black dots represent
leading order interactions. Crossed diagram not shown.
Right panel: The real part of the electric dipole amplitude $E_{0+}$ in the
threshold region. $E_\gamma$ is the photon energy in the lab frame. The data
are from Ref.~\cite{Schmidt:2001vg} and the fourth order baryon CHPT calculation
is from Ref.~\cite{Bernard:2005dj}.}  
\label{fig:E0p}
\vspace{-3mm}
\end{figure}

Consider the reaction $\gamma p\to \pi^0 p$ in the threshold region, where the
pion three-momentum is very small, $\vec{q}_\pi \simeq 0$. At threshold, a low-energy theorem (LET)
had been derived in~\cite{DeBaenst:1970dqx,Vainshtein:1972ih},
\begin{equation}
\label{eq:LEG}  
E_{0+, \rm thr}= -\frac{e g_{\pi N}}{8\pi m_p} \left[\mu
- \frac{1}{2} (3+\kappa_p) \mu^2\right] = -2.3 \cdot \frac{10^{-3}}{M_\pi}
~ , ~~\mu = \frac{M_\pi}{m_p}\simeq \frac{1}{7}~,
\end{equation}
in terms of the pion-nucleon coupling constant $ g_{\pi N}$, the proton anomalous
magnetic moment, $\mu_p$, the proton mass $m_p$ and the pion mass $M_\pi$. 
This LET was challenged by the measurements at Saclay~\cite{Mazzucato:1986dz}
and Mainz~\cite{Beck:1990da}, which gave much smaller values for $E_{0+, \rm thr}$ than given
in Eq.~(\ref{eq:LEG}). A flurry of theoretical papers was published to resurrect the LET. This is
were Aron entered the scene. In the paper ``Threshold pion
photoproduction and chiral invariance,'' co-authored with Barry Holstein, he argued
that the Saclay and Mainz measurements were subject to corrections that brought the
results in agreement with the LET \cite{Bernstein:1991uj} -- ``all roads lead to Rome'',
as they stated. But when it comes to chiral dynamics, all ways indeed lead to Berne!
In 1991, V\'eronique Bernard, J\"urg Gasser, Norbert Kaiser and I (BGKM) reanalyzed the LET in
baryon chiral perturbation theory (CHPT), and found that the Taylor expansion in the
energy is not well behaved, that is the so-called triangle diagram, see the left panel
of Fig.~\ref{fig:E0p}, starts to contribute already at ${\cal O}(\mu^2)$ and not at
${\cal O}(\mu^3)$ as expected from the power counting for
one-loop diagrams. In fact, the authors of Ref.~\cite{Vainshtein:1972ih} were aware of
such IR effects (non-analyticities in the amplitudes) but did not consider the triangle
diagram. So the LET really reads (in fact, the old and incorrect LET was later called a
LEG (low-energy guess)~\cite{Ecker:1994ra})
\begin{equation}
\label{eq:LET}  
E_{0+,\rm thr} = -\frac{e g_{\pi N}}{8\pi m_p}
\left[ \mu - \left(\frac{3+\kappa_p}{2} + \frac{m_p^2}{16F_\pi^2}\right) \mu^2
+ {\cal O}(\mu^3)\right]~,
\end{equation}
where the new term at  ${\cal O}(\mu^2)$ is numerically sizable and the issue of
convergence of the series arises. Thus, one needs to work out the ${\cal O}(\mu^3)$
corrections, which was achieved by BKM in 1994~\cite{Bernard:1994gm}. In fact, to
verify the proper LET, progress was made in two strongly intertwined strands. On the
theory side, BKM were improving the theory and the fits to the Mainz data \cite{Bernard:2001gz},
whereas on the experimental side, Reinhard Beck, his students and Aron were performing improved
experiments~\cite{Fuchs:1996ja}. Note also the experimental activity at
Saskatoon~\cite{Bergstrom:1996fq,Bergstrom:1997jc}.
This culminated on the experimental side with the 2001
paper by Schmidt et al. \cite{Schmidt:2001vg} and on the theory side with the improved fourth
order calculation by Bernard, Kubis and myself~\cite{Bernard:2005dj}, see the right panel
of Fig.~\ref{fig:E0p}. As we will see later, this is  still state-of-the-art in the determination
of $E_{0+,\rm thr}$. More details on this intriguing story are given in the review~\cite{Bernard:2006gx}.
It is also worth noticing in this context that new P-wave LETs~\cite{Bernard:1994gm,Bernard:2001gz}
were successfully tested, but this is a different story not told here.

\begin{figure}[t!]
\begin{center}
\includegraphics[width=0.35\textwidth]{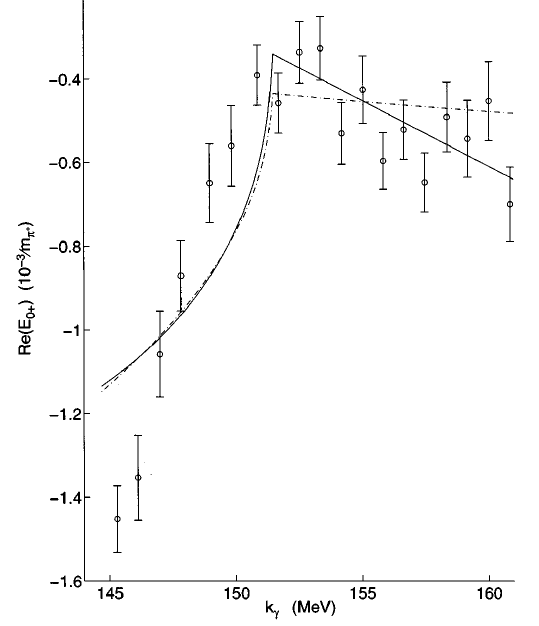}~~
\includegraphics[width=0.55\textwidth]{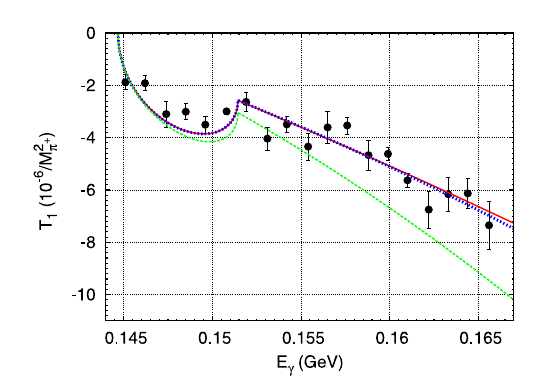}~~
\caption{Left panel: Re~$E_{0+}$ versus the photon energy. The circles represent the multipole fit,
the solid line represents the unitary fit based on Eq.~(\ref{eq:cusp}) and the dashed-dotted line is the
fit based on CHPT. Figure from Ref.~\cite{Bernstein:1996vf}.
Right panel: $T_1$ extracted from the data of Ref.~\cite{Schmidt:2001vg} using the following approaches:
SP (dotted), SPD (solid) and SP with the D-waves added without refitting the data (dashed). Figure
from Ref.~\cite{Fernandez-Ramirez:2009zmg}}.  
\label{fig:cusp}
\end{center}
\vspace{-5mm}
\end{figure}

In fact, Aron was even more interested in the unitary cusp due to the opening of the $\pi^+n$ threshold
just $6.76$~MeV above the $\pi^0p$ threshold. Cusps had been predicted by Wigner already in 1948
\cite{Wigner:1948zz} but had proven to be elusive in experiment, and are still a hot topic nowadays, 
see e.g.~\cite{Guo:2019twa}. In the context of neutral
pion photoproduction, a cusp was predicted by H\"ohler and M\"ullensiefen in 1959~\cite{HoeMue:1959},
\begin{equation}
\label{eq:cusp}
E_{0+}^{\gamma p\to \pi^0p}(k_\gamma) = A_0(k_\gamma) + i \beta \, q_+~,~~\beta
= E_{0+}^{\gamma p\to \pi^+ n} \cdot a_{\rm CEX}^{\pi^+ n \leftrightarrow \pi^0 p}~,
\end{equation}  
with $q_+$ the momentum of the charged pion in the $\pi^+n$-channel and
the strength of the cusp $\beta$ featuring the pion-nucleon charge exchange (CEX) scattering
length $a_{\rm CEX}^{\pi^+ n \leftrightarrow \pi^0 p}$. Further, $ E_{0+}^{\gamma p\to \pi^+ n}$ is given by
the famous Kroll-Ruderman term. This formula was rediscovered in the 1980ties
by F\"aldt~\cite{Faldt:1979fs}, Laget~\cite{Laget:1981jq}
and others. Aron realized early the importance of measuring $\beta$ to
get a handle on this $\pi N$ scattering length as a test of chiral dynamics, see his
talk at the Blaubeuren meeting in 1995~\cite{Bernstein:1995yw}. In the paper ``Observation of a unitary cusp in 
the threshold $\gamma p\to \pi^0 p$ reaction''~\cite{Bernstein:1996vf} Aron and coworkers performed a rigorous
multipole analysis of the threshold data from MAMI and indeed observed the unitary cusp, see the left
panel of Fig.~\ref{fig:cusp}. Clearly, from these data any precision determination of the thought-after
$\pi N$ scattering length was not possible, but in this paper it was also shown how the measurement
of polarization observables in $\vec{\gamma} p\to \pi^0p$ would make such a determination possible.
For an update on the prediction for the cusp, see Ref.~\cite{Hoferichter:2009ez}.

From the beginning of his considerations of neutral  pion photoproduction, Aron was always intrigued to
get a handle on isospin-breaking effects. This was on the one hand inspired by Steve Weinberg's 1977 paper
on the ``Problem of Mass'', where he predicted a huge isospin breaking effect in the scattering lengths for neutral pion
scattering off protons and neutrons, $a(\pi^0 p) = 1.4 \cdot 10^{-15}$cm versus  $a(\pi^0 n) = 1.9 \cdot 10^{-15}$cm, 
e.g. these differ by more than 30\% while isospin conservation would say that they should be equal~\cite{Weinberg:1977hb}.
One the other hand, Aron was fascinated by the work of J\"urg Gasser and Heiri Leutwyler on the determination
of the light quark masses~\cite{Gasser:1982ap,Leutwyler:1996qg}, a topic closely related to strong isospin breaking.
In the paper ``Light quark mass difference and isospin breaking in electromagnetic pion production'' \cite{Bernstein:1998ip},
he developed a three-channel generalization of the Fermi-Watson theorem to relate isospin breaking in photopion
production and $\pi N$ scattering. In particular, he proposed to measure the target asymmetry $T$ to determine
isospin violation in the cusp strength $\beta$ respectively in $a_{\rm CEX}^{\pi^+ n \leftrightarrow \pi^0 p}$.
Furthermore, this paper contains a proposal to measure  $a(\pi^0 p)$ via the imaginary part of the $E_{0+}$ amplitude
below the $\pi^+ n$ threshold, a very daunting enterprise but a first step to check Weinberg's prediction.
I consider this paper Aron's photoproduction masterpiece. His view on the status of pion photoproduction and the
opportunities at MAMI and HI$\gamma$S is
nicely summarized in the 2009 review article~\cite{Bernstein:2009dc}.

However, Aron was still not completely satisfied with the theory of threshold pion photoproduction. With
Cesar Fern\'andez-Ram\'irez and Bill Donnelly he investigated the effects of the D-waves in neutral pion photoproduction
\cite{Fernandez-Ramirez:2009zmg,Fernandez-Ramirez:2009wge}. As usual, one expands the differential cross section
in the threshold region in terms of Legendre polynomials $P_i(\theta)$,
\begin{equation}
\frac{d\sigma}{d\Omega} \propto T_0 + T_1 P_1(\theta) + T_2  P_2(\theta) +  T_3  P_3(\theta) +  T_4  P_4(\theta) + \ldots ,
\end{equation}  
where the $T_i$ depend on the energy. If one takes the S- and P-waves into account, then only the $T_{0,1,2}$ contribute
(these are often called $A,B,C$), whereas adding the D-waves leads to the appearance of the the quantities $T_3$
and $T_4$. For the $T_1$-coefficient, this amounts to
\begin{equation}
T_1 =  2\,{\rm Re}~\left[E_{0+}^{} P_1^* \right] + \delta T_1~,
\end{equation}
with $P_1 = 3E_{1+}+M_{1+}-M_{1-}$ the conventional combination of electric and magnetic P-waves and $\delta T_1$
emerges due to the P/D-wave interference. This effect is particularly pronounced above the $\pi^+n$ threshold as
shown in the right panel of Fig.~\ref{fig:cusp}. So the inclusion of the D-waves improves the accuracy and their
effect is most visibly seen in the polarization observables.

\begin{figure}[t!]
\begin{center}
\includegraphics[width=0.48\textwidth]{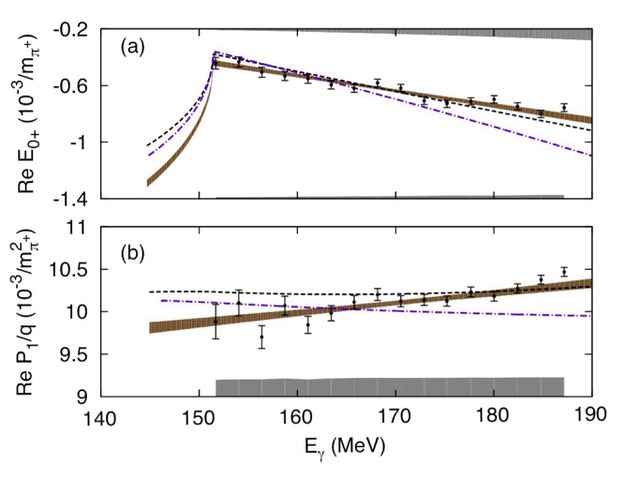}
\includegraphics[width=0.48\textwidth]{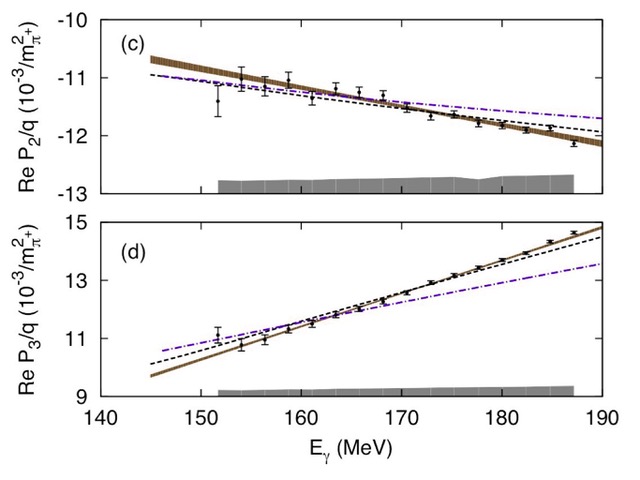}
\end{center}
\vspace*{-3mm}
\caption{S- and P-wave multipoles versus the photon energy in the lab frame.
Left panel: (a) Re~$E_{0+}$.  (b) Re~$P_1/q$. Right panel: (c) Re~$P_2/q$. (d) Re~$P_3/q$.
The data are depicted by the black circles with error bars. The green bands represent the empirical fits based on the 
unitary SP approximation. The dashed black line denotes the heavy baryon CHPT result and the blue dotted line
is based on covariant baryon CHPT~\cite{Hilt:2013uf}.
Figure courtesy of Dave Hornidge.} 
\label{fig:dave}
\vspace{-5mm}
\end{figure}

All these efforts resulted in the MAMI proposal A2-10/09  with the title ``Measurement of Polarized Target and Beam 
Asymmetries in Pion-Photoproduction on the Proton: Test of Chiral Dynamics'' of the A2 collaboration,
led by Michael Ostrick, Dave Hornidge,
Wouter Deconinck and Aron, which was quite ambitious, proposing precise measurements of $\vec{\gamma} p \to \pi^0 p$ 
from threshold up to the region of the $\Delta$-resonance, with the aim of providing stringent tests of the chiral dynamics
of QCD. After lots of work this resulted in the paper ``Accurate Test of Chiral Dynamics in the $\vec{\gamma} p \to \pi^0 p$ 
Reaction''~\cite{A2:2012lnr}. In that paper, precision measurements of the differential cross sections $d\sigma/d\Omega$
and the linearly polarized photon asymmetry $\Sigma = (d\sigma_\perp - d\sigma_{\Vert})/ (d\sigma_\perp - d\sigma_{\Vert})$ were
reported. In fact, the energy dependence of the photon asymmetry $\Sigma$ had never been measured, e.g. in
Ref.~\cite{Schmidt:2001vg} this quantity was averaged and given for just one energy.
These data allowed for a precise determination of the energy dependence of the real parts of the S- and all 
three P-wave amplitudes for the first time and provided the most stringent test of the predictions of CHPT
and its energy region of agreement with experiment, see Fig.~\ref{fig:dave}. There is one drawback with these data,
as a close inspection of Fig.~\ref{fig:dave} reveals: Statistically significant data could only be taken starting
from the $\pi^+ n$ threshold, that is why the energy dependence of Re~$E_{0+}$ shown in Fig.~\ref{fig:E0p} still
defines the state-of-the-art from the $\pi^0 p$ threshold up to the cusp. The reason for this was the increase
in energy of the MAMI accelerator, which benefited many experiments but not the extraction of the multipoles
at and very close to the threshold.

\begin{figure}[t]
\begin{center}
\includegraphics[width=0.50\textwidth]{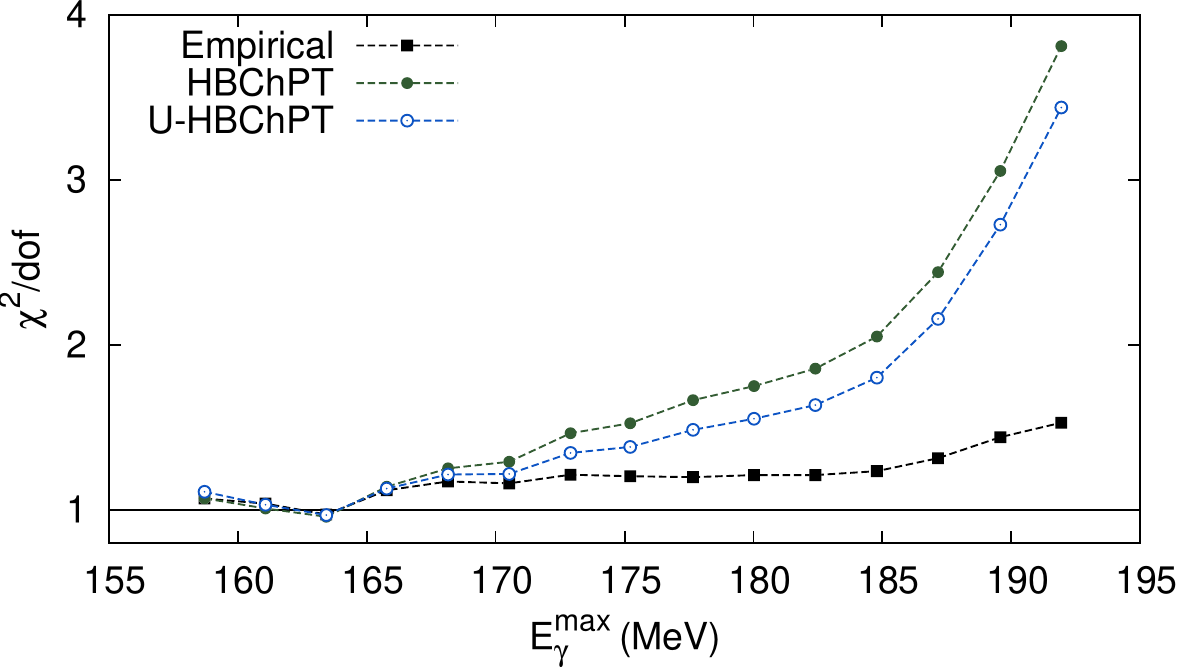}
\caption{$\chi^2$/dof energy dependence fits from the photon energy $E_\gamma^{\rm min} = 151.7\,$MeV
up to the variable maximum energy $E_\gamma^{\rm max}$. The black squares, open blue circles and full
green circles represent the empirical, the HBCHPT and the U-HBCHPT fit, in order. 
  The lines are drawn to guide the eye.  
Figure adapted from Ref.~\cite{Fernandez-Ramirez:2012vlt}.}.  
\label{fig:chi2}
\end{center}
\vspace{-7mm}
\end{figure}

One last issue in neutral pion photoproduction concerns the range of validity of chiral perturbation theory.
From the beginning of our discussions on this topic, Aron always asked me how far above threshold one could
go with heavy baron CHPT? My answer, which was largely based on intuition, always was that $E_\gamma^{\rm max} =
170\,$MeV would be the limit, may be $180$~MeV if one is optimistic. Remember that the threshold is located
at $E_\gamma^{\rm thr} = 144.7\,$MeV. Note also that in all our papers, BKM and associates showed the multipoles
up to $E_\gamma^{\rm max} =165\,$MeV, see e.g.~\cite{Bernard:1994gm,Bernard:2001gz,Bernard:2005dj}. This question was
finally answered by Aron himself in collaboration with Cesar Fern\'andez-Ram\'irez. In the paper
``Upper Energy Limit of Heavy Baryon Chiral Perturbation Theory in Neutral Pion Photoproduction''
\cite{Fernandez-Ramirez:2012vlt}, they compared HBCHPT at ${\cal O}(p^4)$ \cite{Bernard:1994gm,Bernard:2001gz}
with a unitarized version (U-HBCHPT), in which Im~$E_{0+}$ was taken from the exact unitarity relation Eq.~\eqref{eq:cusp},
and the S- and P-wave fit to the precise data from Ref.~\cite{A2:2012lnr}. The outcome is shown in Fig.~\ref{fig:chi2}.
There is very good agreement up to  $E_\gamma^{\rm max} = 170\,$MeV and marked deviations start to appear for higher
maximal energies. So my intuition was not that bad and Aron put the answer on firm grounds.

Further progress was made by the A2 collaboration in Ref.~\cite{MAINZ-A2:2015yzu}. They presented
measurements of the polarization-dependent cross sections $\sigma_T$ associated with the target
asymmetry $T$ for the reaction $\gamma\vec{p}\to \pi^0p$ with transverse polarization from the
$\pi^0$ threshold to energies of 190~MeV. This allowed for the first time for a direct extraction
of the imaginary part of the $E_{0+}$ multipole. Within uncertainties, this result is in agreement
with the fourth order HBCHPT prediction as well as relativistic predictions from Refs.~\cite{Hilt:2013uf,Gasparyan:2010xz}.
This story of neutral pion photoproduction can be considered a flagship for the fruitful interplay of
theory and experiment, exactly what always has been on Aron's mind. 
I will come back to this issue at the end of the talk.

A few remarks on neutral  pion electroproduction, that is the reaction $\gamma^\star(Q^2) p\to \pi^0 p$, are in order.
Here, due to the virtuality $Q^2$ of the virtual photon, the kinematics is more complicated as compared to the
real photon case. But this is also a blessing as new longitudinal multipoles arise that allow for further
tests of chiral dynamics,
described in terms of appropriate LETs \cite{Bernard:1992ms,Bernard:1994dt}. The HBCHPT formalism for this reaction was
developed by BKM and Harry Lee in Refs.~\cite{Bernard:1993bq,Bernard:1996bi}.
Early experiments from NIKHEF~\cite{vandenBrink:1995uka,vandenBrink:1997cs} and Mainz~\cite{Distler:1998ae,Merkel:2001qg}
showed some agreements but also disagreements with the HBCHPT predictions. Aron and I were involved in the Bigbite collaboration
proposal on  ``Precision Measurements of Electroproduction of $\pi^0$ near threshold: A Test of Chiral QCD Dynamics''
at Jefferson Lab~\cite{Bigbite} , which intended to settle all these questions. The experiment took some time to finally produce
interesting data that were published in 2015, co-authored by Aron~\cite{HallA:2015gcv}. The outcome of this work is that
the S-wave cross section $a_0$ is in reasonable agreement with the CHPT predictions, but the P-wave coefficient $b$ in the
threshold cross section $A_0^{T+L}= a_0 + b |\vec{p}_\pi|^2$, see Ref.~\cite{HallA:2015gcv} for details,
starts to deviate from the predictions already at low virtualities $Q^2 \simeq 0.07\,$GeV$^2$. Clearly,
more experimental and theoretical work in this field is needed.

\section{The chiral anomaly and the neutral pion lifetime}
\label{sec:anom}

\begin{figure}[t]
\begin{center}
\includegraphics[trim=0 -15ex 0 0,width=0.45\textwidth]{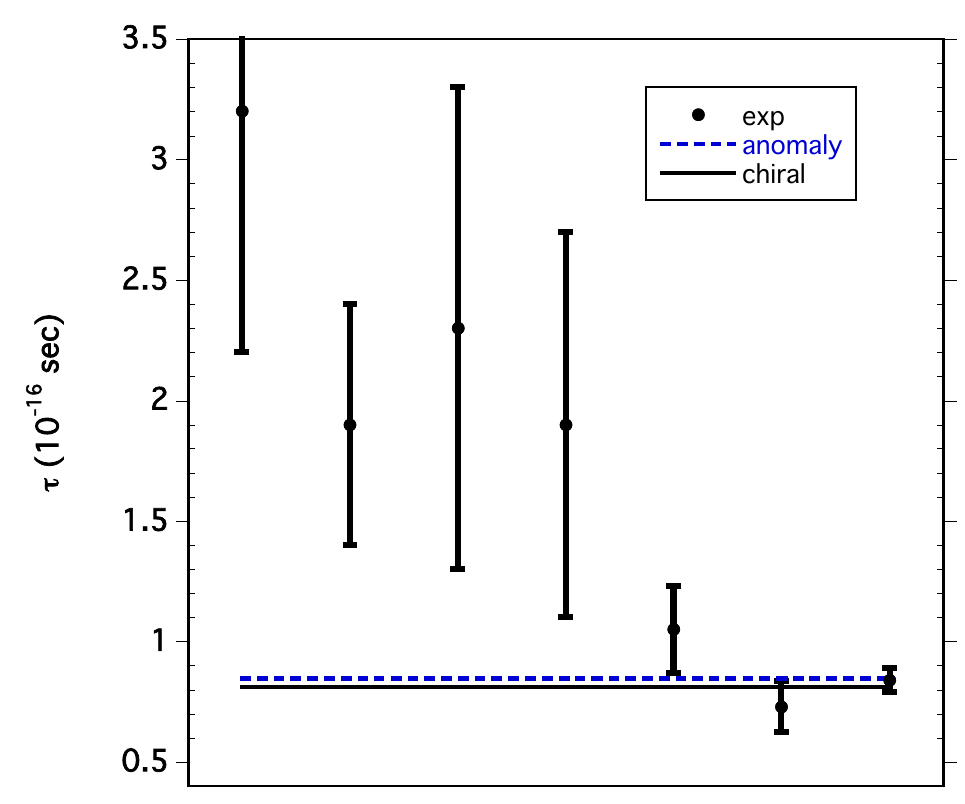}~~
\includegraphics[width=0.45\textwidth]{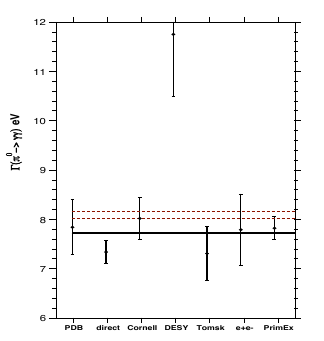}
\caption{Left panel: Early measurements of the neutral pion lifetime in comparison to the anomaly prediction
and the leading chiral corrections due to pseudoscalar meson mixing.  Figure courtesy of Aron Bernstein.
Right panel: Preliminary PrimEx results presented by Aron in his 2009 Chiral Dynamics  talk.
Figure from Ref.~\cite{Bernstein:2009zz}}.  
\label{fig:pi02g}
\end{center}
\vspace{-7mm}
\end{figure}

As it is well known, the decay $\pi^0\to 2 \gamma$ has revealed  anomalous symmetry breaking, meaning
that  quantum corrections break a symmetry of the classical theory, see the work by Adler \cite{Adler:1969gk},
Bell and Jackiw~\cite{Bell:1969ts} in 1969 and many others to follow. Calculating the pertinent VVA triangle
diagram and its crossed partner, one obtains a precise prediction at leading order for the width (lifetime) of
the neutral pion
\begin{equation}
\Gamma^{\rm anom}_{\pi^0\gamma\gamma}=\frac{\alpha^2 M_\pi^3}{64\pi^3 F_\pi^2} = 7.76~{\rm eV}~.
\end{equation}
Early measurements were rather inconclusive, related to the fact that this lifetime is difficult to measure,
see the left panel of Fig.~\ref{fig:pi02g} and the detailed discussion in the nice review by Aron and Barry 
Holstein~\cite{Bernstein:2011bx}. On the theoretical side, the first calculations of corrections to this leading
order result in the framework of chiral perturbation theory based on the Wess-Zumino-Witten Lagrangian were done in Refs.~\cite{Donoghue:1985jkn,Bijnens:1988kx},
leading to a few percent correction mostly due to $\pi^0$-$\eta$-$\eta'$ mixing.  Aron was intrigued by Bachir Moussallam's 
work on $\pi^0, \eta, \eta'\to 2\gamma$ decays in CHPT when  he had a sabbatical  at MIT and worked on chiral
sum rules and higher-order corrections to $\pi^0$, $\eta$ and $\eta'$ decays~\cite{Moussallam:1994xp}. Based on these
observations and early works, Aron was one of the initiators of the PrimEx proposal~\cite{PrimEx}. This experiment
intended to measure the neutral pion lifetime with an error of 1.5\%, a very significant improvement in accuracy compared
to earlier measurements. Consequently, the race was on to improve the theoretical predictions based on CHPT with
dynamical photons or CHPT combined with the $1/N_c$ expansion, which led to two
publications which appeared within one month in 2002, on the one side the work by Balasubramanian Ananthanarayan and
Bachir Moussallam (AM)~\cite{Ananthanarayan:2002kj} and on the other side the one of Jos\'e Goity, Aron and
Barry Holstein (GBH)~\cite{Goity:2002nn}, with very similar results
\begin{equation}
\Gamma_{\pi^0\gamma\gamma}= \left\{ 
\begin{matrix}8.06\pm 0.02\pm 0.06~{\rm eV}~(\rm {AM})~,\\
 8.10 \pm 0.08~{\rm eV}~~~~~~~~~({\rm GBH})~. \end{matrix}\right.
\end{equation}
This was later improved to two-loop accuracy in Ref.~\cite{Kampf:2009tk} with the result $\Gamma_{\pi^0\gamma\gamma}
= 8.09 \pm 0.11$~eV. For more details on these theoretical developments, I refer to Kampf's talk at this workshop~\cite{Kampf}.

Aron presented preliminary results of the PrimEx experiment at the 6th Chiral Dynamics meeting in Bern in 2009,
see the right panel of Fig.~\ref{fig:pi02g} and the corresponding PrimEx paper was published in 2011~\cite{PrimEx:2010fvg}.
It was high-lighted in Aron's opening talk at the 7th Chiral Dynamics meeting in 2012 at Jefferson
Laboratory~\cite{Bernstein:2013aia}. The measured value was
\begin{equation}
\Gamma_{\pi^0\gamma\gamma}= 7.82\pm 0.14\pm 0.17~{\rm eV}~,
\end{equation}
so the achieved accuracy was 2.8\%, somewhat above the targeted value. After this groundbreaking
result, Aron teamed up again with Barry Holstein to write the nice review on  ``Neutral pion lifetime
measurements and the QCD chiral anomaly,'' published in {\em Review of Modern Physics} in 2013~\cite{Bernstein:2011bx}.
This is  another testimony of Aron's legacy! The 1.5\% accuracy for the pion lifetime measurement was finally
achieved by the the PrimEx-II experiment, which reported $\Gamma_{\pi^0\gamma\gamma}= 7.80\pm 0.06\pm 0.11$~eV
in 2020~\cite{PrimEx-II:2020jwd}. More details on this achievement are given in Gasparian's talk at this
workshop~\cite{Gasparian}.

Aron was also interested in the decays of pseudoscalar mesons ($PS$), $PS\to \gamma\gamma^\star$, and the radii of
the $PS$ mesons. In his
contribution to the 8th Chiral Dynamics meeting in Pisa, he derived the radii of pseudoscalar mesons the
decays $PS \to \gamma^\star(Q^2) \gamma$ at low $Q^2$,~\cite{Bernstein:2015sac}
\begin{equation}
F(Q^2)=F_{PS}(0)\left(1- Q^2 \langle \frac{1}{6}r^2\rangle + \ldots\right) ~,~~~ F_{PS}(0) = \left(\frac{4\Gamma(PS\to 2\gamma)}
{\pi M_{PS}^3 \alpha^2}\right)^{1/2}~.
\end{equation}
His intention was to stimulate new, accurate experiments and further calculations based on CHPT and lattice QCD.
Note that these transition form factors gained prominence in the theoretical analysis of the muon anomalous magnetic
moment.

\section{Summary and an experimental challenge}

The imprints Aron left in the field of chiral dynamics are best summarized by the talks he gave
at the Chiral Dynamics meetings over more than two decades, as summarized in Table~\ref{tab:AronCD}.
These cover the issues discussed here plus other topics in chiral dynamics, that were less central
to Aron's work. Also, he was one of the founding fathers of this successful workshop series and over
the years enthusiastically helped to shape the program of each single meeting.
\begin{table}[h]
\begin{center}  
\begin{tabular}{|c|c|c|c|}
\hline    
Year & Place & Proc. & Title of Aron's talk\\
\hline
1994 & MIT & \cite{Bernstein:1995vvb}      & none - main organizer\\
1997 & Mainz & \cite{Bernstein:1998pm}     & {\em Introduction to Chiral Dynamics: Theory and Experiment}\\
2000 & JLab & \cite{Bernstein:2001rk}      & {\em Experimental Chiral Dynamics}\\
2003 & Bonn & \cite{Meissner:2003hr}      & {\em Hadron Deformation and Chiral Dynamics}\\
2006 & Duke & \cite{CD2006}      & {\em Opening Remarks: Experimental Tests of Chiral Symmetry Breaking}\\
2009 & Bern & \cite{CD2009}      & {\em Lifetime Measurement of the $\pi^0$ Meson and the QCD Chiral Anomaly}\\
2012 & JLab & \cite{CD2012}      & {\em Outlook} \\
2015 & Pisa & \cite{CD2015}      & {\em The $\pi^0,\eta,\eta'\to \gamma\gamma^\star$ Decay Rates and Radii}\\
2018 & Duke &  \cite{CD2018}     & none - could not attend, but very active organizer\\
\hline
\end{tabular}
\vspace*{-2mm}
\end{center}
\caption{Summary of the first to ninth ``Chiral Dynamics - Theory \& Experiment'' workshop with its
  year, location, the reference to the proceedings and Aron's talk or role.\label{tab:AronCD}}
\end{table}

\begin{figure}[t]
\begin{center}
\includegraphics[trim=0 -15ex 0 0,width=0.40\textwidth]{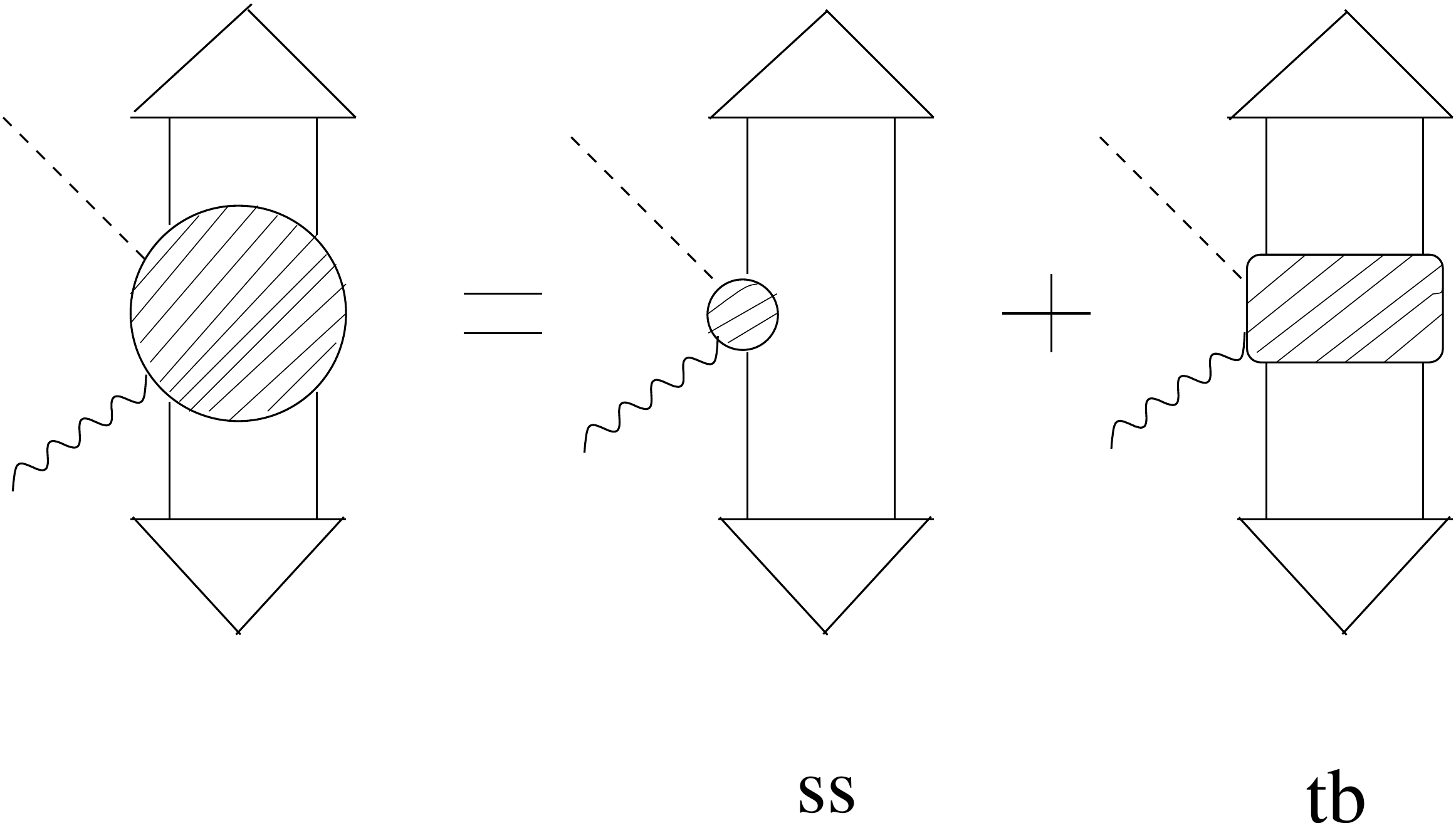}~~~~~~~~~~
\includegraphics[width=0.45\textwidth]{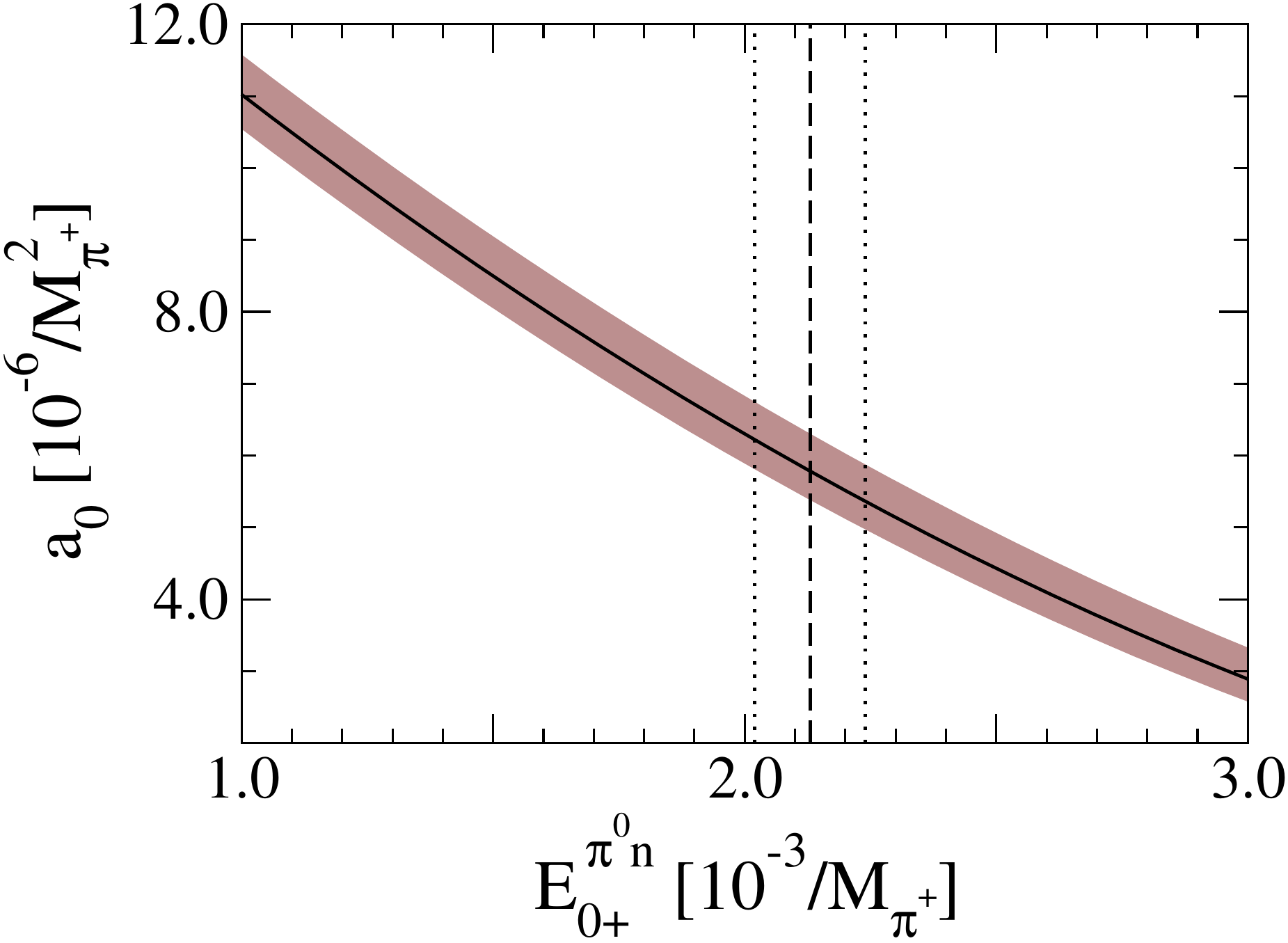}
\caption{Left panel: Feynman diagrams for $\gamma d\to \pi^0 d$. Solid, dashed and wiggly lines denote nucleons, pions and
photons, in order. The triangle represents the deuteron wave function.
The single scattering (ss) amplitude is sensitive to pion photoproduction on the
proton and the neutron, whereas the three-body (tb) amplitude subsumes all corrections from meson-exchange
currents in terms of pion exchanges and local four-nucleon vertices.
Right panel: Sensitivity of the S-wave cross section $a_0$ to the single-neutron multipole $E_{0+}^{\pi^0 n}$
in $\pi^0$ production off $^3$He.
The vertical dashed line gives the CHPT prediction for $E_{0+}^{\pi^0 n}$ and the vertical dotted lines indicate
a 5\% error in this quantity. The shaded band gives an estimate of the theory error from the few-body calculation
combined with the one in the neutron amplitude. Figure from Ref.~\cite{Lenkewitz:2012ixw}}.  
\label{fig:pi0-n}
\end{center}
\vspace{-7mm}
\end{figure}

In Aron's spirit, I will end this contribution with a challenge to the experimentalists. To be specific,
consider now the reaction $\gamma n\to \pi^0 n$ in the threshold region. In the classical dipole picture,
we have $E_{0+}(\gamma n \to \pi^0 n) = 0$~\cite{Ericson:1988gk}. In CHPT, one obtains the
counter-intuitive prediction~\cite{Bernard:1994gm}
\begin{equation}
E_{0+}^{\pi^0 n} = 2.1 > |E_{0+}^{\pi^0 p}| \simeq 1.2~,
\end{equation}  
in the canonical units. This is truly remarkable as $|E_{0+}^{\pi^0 n}| \simeq 2|E_{0+}^{\pi^0 p}|$ shows that
quantum effects clearly defy intuition. Since free neutron targets are not available, a first test of
the prediction can be made on the deuteron, which is an isoscalar target and the S-wave amplitude for
neutral pion photoproudction takes the form
\begin{equation}
E_d^{} = E_d^{\rm ss} + E_d^{\rm tb}~, \quad E_d^{\rm ss} \propto E_{0+}^{\pi^0p} + E_{0+}^{\pi^0n}~,
\end{equation}  
where ``ss'' denotes the single scattering amplitude and ``tb'' are the three-body corrections
in Weinberg's notation~\cite{Weinberg:1992yk}, see also the left panel of Fig.~\ref{fig:pi0-n}.
Such a few-body calculation was performed by Silas Beane, V\'eronique Bernard, Harry Lee, Bira van Kolck and me.
In a hybrid calculation with precise wave functions from semi-phenomenological models and amplitudes
from HBCHPT, in Ref.~\cite{Beane:1997iv}  it was shown that the three-body terms are dominant but converge quickly,
leading to the prediction $E_d^{\rm th} = -1.8 \pm 0.2$, based on the CHPT predictions for $E_{0+}^{\pi^0 p}$ and
$E_{0+}^{\pi^0 n}$. This is consistent with the old Saclay data that were reanalyzed in 1987,
leading to $E_d^{\rm exp} = -1.7 \pm 0.2$~\cite{Argan:1987dm}. However, both values are
presumably afflicted with larger uncertainties. This was later extended to neutral pion electroproduction
off the deuteron in Refs.~\cite{Bernard:1999ff,Krebs:2004ir}.
The topic of extracting the S-wave multipole $E_{0+}^{\pi^0 n}$ in photoproduction off light nuclei
was taken up by Mark Lenkewitz, Evgeny Epelbaum, Hans-Werner Hammer and me in Refs.~\cite{Lenkewitz:2011jd,Lenkewitz:2012ixw}.
There, the sensitivity of the threshold
cross section $a_0$ for $\gamma + ^3{\rm H} \to \pi^0 + ^3{\rm H}$ to the neutron S-wave amplitude was worked
out with operators and wave functions from chiral EFT. The S-wave cross section is defined as
\begin{equation}
a_0 = \frac{|\vec{k}_\gamma|}{|\vec{q}_\pi|}\frac{d\sigma}{d\Omega}\biggr|_{\vec{q}_\pi=0} = \left|E_{0+}\right|^2~,
\end{equation}   
with $\vec{k}_\gamma$ and $\vec{q}_\pi$ the photon and the pion three-momentum, respectively. The resulting
cross section $a_0$ as as function of the elementary  neutron S-wave amplitude is shown in the right panel
of Fig.~~\ref{fig:pi0-n}, indicating that $^3$He is a promising candidate to test the CHPT prediction for
$E_{0+}^{\pi^0n}$. Therefore, such an experiment should be performed -- a true challenge for my experimental colleagues.
Aron would have loved to see/perform such an experiment.

I hope that with this contribution I could express my deep appreciation for Aron's work in the field of
chiral dynamics, with this meeting serving as yet one more testimony. On the personal side, let me finish
by noting that  over the years, Aron has not only been an esteemed  colleague but also has become a good friend.

\section*{Acknowledgments}
I am deeply grateful to Aron for all the fun we had doing physics and to Susan and Aron for
providing good food and pleasant times at their home. Many thanks also to Reinhard Beck and
Dave Hornidge for their help in providing material and, of course, all my collaborators who helped to shape
my understanding of the issues discussed here. The work  is supported in
part by the Deutsche Forschungsgemeinschaft (DFG, German Research Foundation) and the
NSFC through the funds provided to the Sino-German Collaborative  
Research Center TRR~110 ``Symmetries and the Emergence of Structure in QCD''
(DFG Project-ID 196253076 - TRR 110, NSFC Grant No. 12070131001),
by the Chinese Aca\-de\-my of Sciences (CAS) through a President's
International Fellowship Initiative (PIFI) (Grant No. 2018DM0034)
and by the VolkswagenStiftung (Grant No. 93562).

\end{document}